\documentclass[reprint,
groupedaddress,
superscriptaddress,
amsmath,amssymb,
aps,
]{revtex4-2}

\usepackage{graphicx}
\usepackage{dcolumn}
\usepackage{bm}
\usepackage{lineno}
\usepackage{graphicx}
\usepackage{dcolumn}
\usepackage{bm}
\usepackage{hyperref}
\usepackage{braket}
\usepackage{lipsum}
\usepackage{comment}
\usepackage{physics}
\usepackage[utf8]{inputenc}
\usepackage{xcolor}

\begin{document}

\title{The rise and fall of patterns in driven-dissipative Rydberg polaritons}

\author{Hadiseh Alaeian}
\email{halaeian@purdue.edu}
\affiliation{Elmore Family School of Electrical and Computer Engineering, Purdue University, West Lafayette, IN 47906, US}
\affiliation{Department of Physics and Astronomy, Purdue University, West Lafayette, IN 47906, US}

\author{Valentin Walther}
\email{vwalther@purdue.edu}
\affiliation{Department of Chemistry, Purdue University, West Lafayette, IN 47906, US}
\affiliation{Department of Physics and Astronomy, Purdue University, West Lafayette, IN 47906, US}


\begin{abstract}
Spatial structures commonly emerge in interacting nonlinear systems. In this study, we focus on the out-of-equilibrium dynamics of the recently-established platform of Rydberg exciton-polaritons, fueled by their characteristic long-range non-local interactions, in the presence of an external drive and dissipation. Our work elucidates how modulational instability sets off spontaneous density pattern formations in a Rydberg polariton system with characteristic scales in the micron range. 
Under conditions of an unstable flattop state, we track the evolution of the polariton ensemble, showing the emergence of meta-stable patterns and their collapse in the long-time limit. We trace this phenomenon back to the destructive interference between the polariton state and the pump in a driven ensemble. Finally, we map out conditions that allow stable patterns to form under incoherent pumping. These findings provide new opportunities for exploring the emerging field of long-range interacting gases through Rydberg exciton-polaritons.
\end{abstract}

\keywords{Driven-dissipative systems, Long-range interaction, Rydberg exciton-polariton, Nonlinear dynamics, Pattern formation}

\maketitle
\emph{Introduction - }Spontaneous symmetry breaking is a pivotal concept in condensed-matter physics. Within this realm, pattern formation emerges as a frequent occurrence in nonlinear systems, marked by the appearance of extended spatial structures. In the past couple of decades, there have been extensive theoretical and experimental explorations of condensed-matter concepts in ultra-cold dipolar atomic gases and Rydberg-dressed Bose-Einstein Condensates (BEC) featuring long-range anisotropic dipole-dipole interactions. 
The long-range nature of dipolar interactions sets them apart from systems characterized by short-range $s$-wave scattering. Noteworthy advancements encompass the observation of the supersolid phase and theoretical predictions of novel phases resulting from quantum fluctuations~\cite{Henkel2010, Henkel2012, Heinonen2019, Hertkorn2021, Zhang2021, ostermann2022, Chomaz2023}, the emergence of roton instabilities in quasi-1D dipolar BECs under the influence of a periodic lattice potential~\cite{Corson2013}, and the study of pattern formation induced by long-range soft-core potentials~\cite{Cinti2014} or through optical feedback in BECs~\cite{Labeyrie2014, Zhang2021A}. As a result, ultra-cold dipolar and Rydberg atomic gases exhibiting long-range interactions have gained prominence as invaluable platforms for scrutinizing quantum many-body physics and phase transitions.

Pattern formation is also ubiquitous in optics, where the interaction between an electromagnetic field and a polarizable medium, coupled with losses, leads to the spontaneous translational symmetry breaking and hence the formation of patterns~\cite{Tlidi94, Ackemann95, ARECCHI99, Zhang2021A}. Over the years, polaritons, with effective photon-photon interactions mediated by excitons, have emerged as a potent platform in nonlinear optics. Their distinct out-of-equilibrium behavior has rendered them a focal point in a plethora of theoretical and experimental investigations into phase transitions such as pattern formations and superfluidity~\cite{Carusotto2004, Carusotto2006, Amo2009, Carusotto2013, Berloff2013}. To date, most of these studies were centered on semiconductor exciton-polaritons characterized by short-range contact interactions. However, the recent observation of highly-excited Rydberg excitons in materials such as perovskite~\cite{Bao2019}, transition metal dichalcogenides~\cite{Biswas2023, Kapuciski2021}, and notably, cuprous oxide (Cu$_2$O)~\cite{Kazimierczuk2014, Amann2020}, along with the successful demonstration of Rydberg exciton-polaritons, has revitalized interest in exploring dipolar and Rydberg physics with 
\emph{Rydberg} polaritons~\cite{Walther2018, Orfanakis2022}.

\begin{figure}
    \centering
    \includegraphics[width=8.3
    cm]{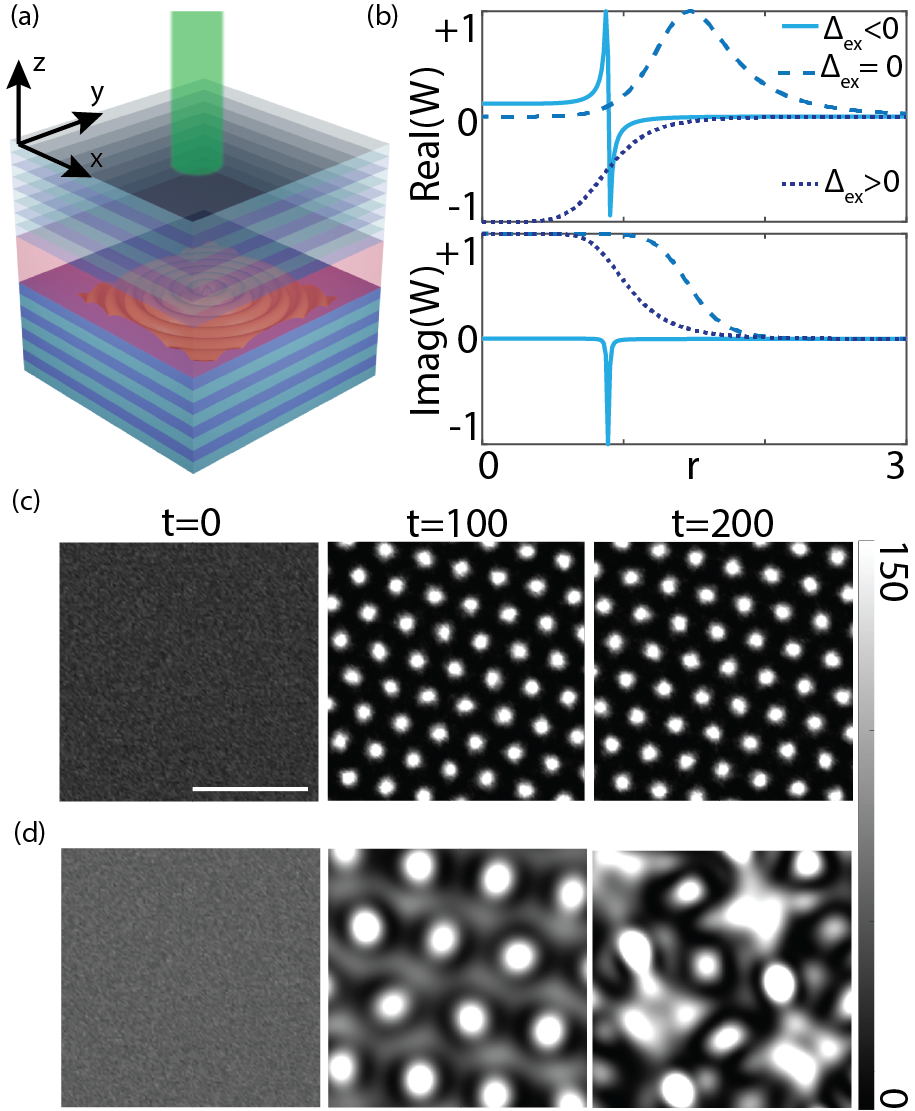}
    \caption{\textbf{Rydberg polaritons in closed and open systems.} (a) Schematics of a 2D optical cavity encapsulating the Rydberg excitonic material which leads to the formation of Rydberg exciton-polaritons. (b) Normalized nonlinear $\chi^{(3)}(r)$ as a function of the transverse coordinate $r$ showing the tunability of dispersive (top) and dissipative (bottom) effective soft-core potential via the exciton detuning, $\Delta_\textrm{ex}$. Snapshots of cavity field intensity in (c) a closed system with $g = 1, C_6 = -1, \Delta = -1$ and (b) an open system with $\Delta_\text{ex} = 6, \gamma = 0.1, \kappa = 0.5, C_6 = -1, \Delta_c = 2$, highlighting the pattern formations in both cases while contrasting their stability. The white scale bar corresponds to $L = 10$.}
    \label{fig:schematics}
\end{figure}

Here, we investigate whether the characteristic traits of long-range interactions in closed systems such as ultra-cold atomic gases persist in out-of-equilibrium Rydberg polaritons. We delve into the dynamic evolution of polariton phases where the homogeneous state becomes unstable, leading to finite-range instability behavior and pattern formation during the system's evolution. Additionally, we scrutinize the long-term stability of this patterned phase in an open system compared to a conservative one, as depicted in Fig.~\ref{fig:schematics}(c),(d). Our calculations indicate that, unlike closed systems, the stability of the driven-dissipative case is contingent on the type of external drive. Given the high tunability of Rydberg polaritons through parameters such as the addressable principal quantum number, cavity parameters, and driving in conjunction with their distinctive long-range interactions, this research offers a unique platform for investigating exotic out-of-equilibrium many-body phases.

Here, we consider a 2D cavity with length $L$ encapsulating the Rydberg excitonic layer hosting Rydberg polaritons, as depicted in Fig.~\ref{fig:schematics}(a). In the mean-field limit of the electromagnetic field in the cavity, the excitons and their correlations can be solved exactly and analytically to the third order in terms of the cavity field which leads to the following generalized Gross-Pitaevskii Equation (GPE) describing the cavity field dynamics only as (cf. Supplementary Materials for more details on the GPE derivation and its range of validity) 
\begin{widetext}
\begin{align}~\label{eq:GPE}
i \partial_{t} \mathcal{E}(\mathbf{r},t) = \left(-\frac{\hbar}{2m_\text{ph}} \nabla^{2} + i\chi^{(1)} - i\frac{\Gamma_{c}}{2} + \int d\mathbf{r'}~W(\mathbf{r} - \mathbf{r'}) \abs{\mathcal{E}(\mathbf{r'},t)}^{2}\right)\mathcal{E}(\mathbf{r},t) + \textrm{drive}\, . 
\end{align}
\end{widetext}
%

\begin{figure*}
    \centering
    \includegraphics[width=18
    cm]{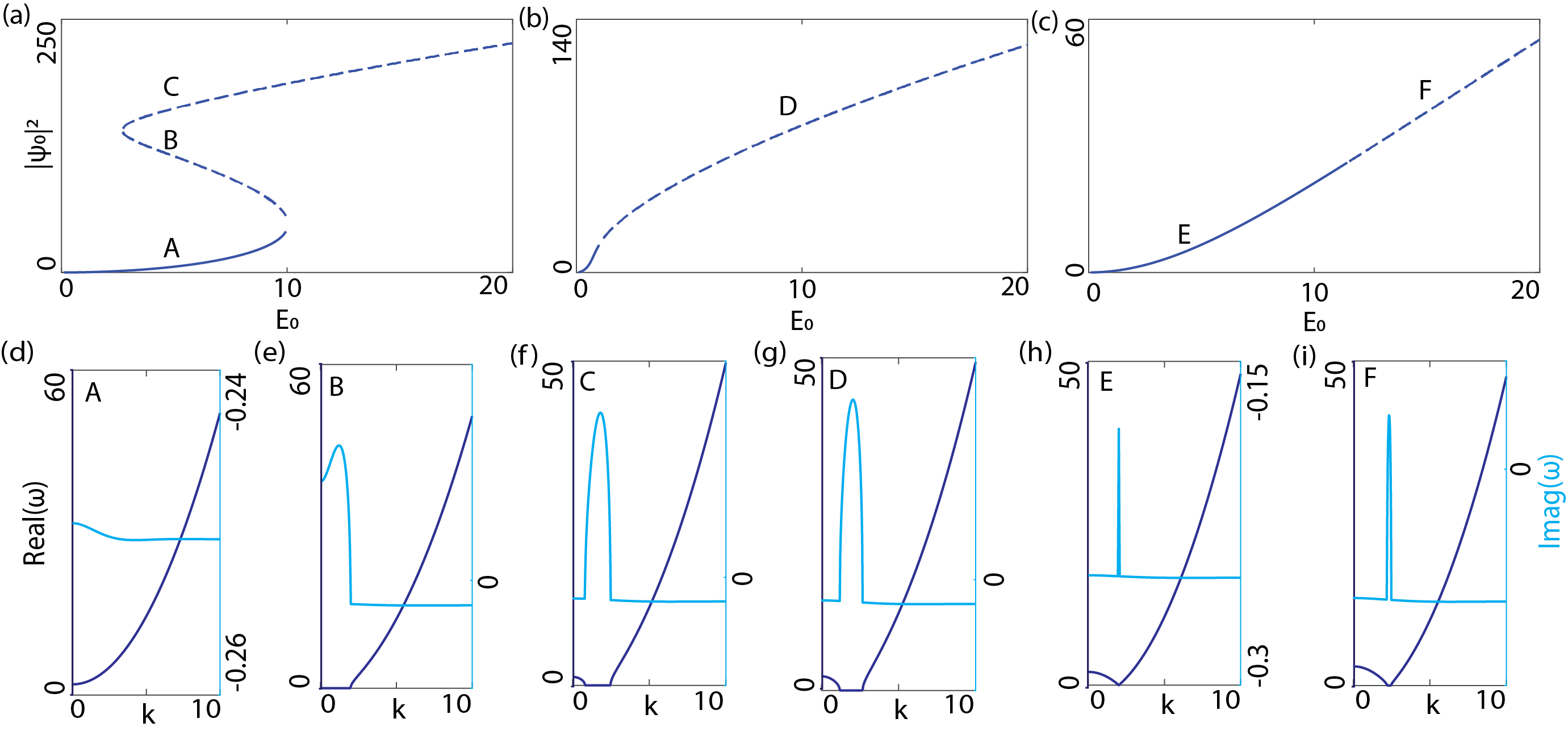}
    \caption{\textbf{Steady states with coherent drive and the Bogoliubov dispersion.} Steady-state photon density ($|\psi_0|^2$) of a Rydberg polariton ensemble, driven by a coherent laser of amplitude $E_0$ where $\Delta_\text{ex} = 6, \gamma = 0.1, \kappa = 0.5, C_6 = -1$, at various cavity detunings (a) $\Delta_c = -2$, (b) $\Delta_c = 0$, and (c) $\Delta_c = 2$. In each panel, stable branches are shown in solid lines while the dashed lines show unstable solutions. (d)-(i) The Bogoliubov dispersion of elementary excitations at points (A)-(F) clarifying the (in)stability of each point. In each panel, the dark (light) blue line shows the real (imaginary) part of $\omega(k)$.}
    \label{fig:SS_cases}
\end{figure*}
%
%
In this equation, $\chi^{(1)}$ and $W$ are the effective linear and nonlinear optical susceptibility, respectively, mediated by the exciton-exciton interaction with the following explicit forms 
\begin{align}~\label{eq:succeptibilities}
    \chi^{(1)} &= -\frac{2 g^{2}}{\Gamma_\text{ex}}\,, \\
    W(\mathbf{r}) &= -\frac{16 g^{4}}{|\Gamma_\text{ex}|^{2}\Gamma_\text{ex}}\frac{ U(\mathbf{r})}{\Gamma_\text{ex} + i U(\mathbf{r})}\, ,
\end{align}
where $U(\mathbf{r})$ is the pairwise long-range van der Waals interaction between excitons as $U(\mathbf{r}) = \frac{C_{6}}{\abs{\mathbf{r} - \mathbf{r'}}^{6}}$ \cite{walther2018interactions}.
The complex-valued exciton and cavity decay rates $\Gamma_\text{ex}$ and $\Gamma_{c}$ are related to the cavity $\Delta_c$ and exciton detuning $\Delta_\textrm{ex}$ as well as the decay rates of the Rydberg state $\gamma$ and cavity photons $\kappa$, via
\begin{align}~\label{eq:generalized decay rates}
    \Delta_c &= \omega_p - \omega_{\text{c}}\,, && \Gamma_c = \kappa - 2 i \Delta_c \, , \\
    \Delta_{\text{ex}} &= \omega_p - \omega_{\text{ex}}\,, && \Gamma_\text{ex} = \gamma - 2 i \Delta_{\text{ex}}\, .
\end{align}
The non-local character of the nonlinear susceptibility stems from the long-range dipole-dipole interactions between excitons, a potential that depends on the principal quantum number through an $n$-dependent $C_6$ coefficient and realizes a dissipative and dispersive soft-core potential, as depicted in Fig.~\ref{fig:schematics}(b)~\cite{Maghrebi15}.
In the rest of the text, we work with the dimensionless quantities through the following scaling
\begin{equation}
   \mathcal{E}(\mathbf{r} , t) \rightarrow \frac{1}{r_0} \mathcal{E}\left(\frac{\mathbf{r}}{r_0},  \frac{t}{\tau}\right)\, ,
\end{equation}
where $c$ is the speed of light, $r_0 = L/(n\pi)$ is the effective cavity length, $n$ is the effective refractive index of the excitonic material, and $\tau = r_0/c$ is the cavity photon round-trip time. Further, in the numerical calculations, we consider $C_6 \leq 0$, corresponding to an attractive interaction.
    
\emph{Results - }Coherent pumping which directly injects photons into the cavity mode can be described via a driving term as $iE_0$ in Eq.~\ref{eq:GPE}. The dynamics of a fluctuation $\delta \mathbf{\Psi}(\mathbf{k},t)$ around any stationary point is then given as 
\begin{equation}
    i\frac{d}{dt}
    \begin{pmatrix}
        \delta \mathbf{\Psi}(\mathbf{k},t)\\
        \delta \mathbf{\Psi}^*(-\mathbf{k},t)
    \end{pmatrix} = \mathcal{B} \begin{pmatrix}
        \delta \mathbf{\Psi}(\mathbf{k},t)\\
        \delta \mathbf{\Psi}^*(-\mathbf{k},t)
    \end{pmatrix}\, ,
\end{equation}
where the Bogoliubov matrix $\mathcal{B}$ for a flattop solution $\psi_0$ is
\begin{widetext}~\label{eq:Bogoliubov open}
    \begin{align}    
    \mathcal{B} = 
       \begin{pmatrix}
          \frac{\mathbf{k}^{2}}{2} + i \left(\chi^{(1)} - \frac{\Gamma_c}{2}\right) +  2\pi |\psi_{0}|^{2}(\Tilde{W}(\mathbf{k}) + \Tilde{W}(0)) & 2 \pi \psi_{0}^{2} \Tilde{W}(\mathbf{k})\\
          -2 \pi \psi_0^{*^2}\Tilde{W}^{*}(\mathbf{k}) & - \frac{\mathbf{k}^{2}}{2} + i \left(\chi^{(1)^*} - \frac{\Gamma_c^*}{2}\right) - 2\pi |\psi_{0}|^{2}(\Tilde{W}^{*}(\mathbf{k}) + \Tilde{W}^*(0))
       \end{pmatrix}\,,
  \end{align}
\end{widetext}
where $\Tilde{W}(\mathbf{k})$ is the Fourier transform of $W(\mathbf{r})$~\footnote{Note that in general the spinor and the Bogoliubov matrix entries must have terms as $\delta \Psi^*(-\mathbf{k},t)$ and $\Tilde{W}(-\mathbf{k})$, i.e. the time-reversed transforms. However, since the functions are centrosymmetric, i.e. $f(\mathbf{r}) = f(-\mathbf{r})$ then the transforms are centrosymmetric as well hence we do the simplification directly.}. This leads to the Bogoliubuv dispersion
\begin{widetext}~\label{eq:open_disp}
    \begin{align}
    \omega_\pm(k) & = i \left(2\pi |\psi_0|^2 \left(\Tilde{W}_I(k) + \Tilde{W}_I(0) \right) + \chi^{(1)}_R - \frac{\kappa}{2} \right) \\ \nonumber
    & \pm \sqrt{\left(\frac{k^2}{2} + 2\pi |\psi_0|^2 \left(\Tilde{W}_R(k) + \Tilde{W}_R(0)\right) - \chi^{(1)}_I - \Delta_c \right)^2 - 4\pi^2 |\psi_0|^4 \left(\Tilde{W}_R^2(k) + \Tilde{W}_I^2(k)\right)}\,,
\end{align}
\end{widetext}
which signifies a modulational instability (MI) where $\text{Im}(\omega(k)) \ge 0$.  

Figure~\ref{fig:SS_cases}(a)-(c) exemplifies the behavior of cavity field intensity for a flattop ansatz $|\psi_0|^2$ as a function of the coherent drive strength for fixed exciton parameters for cavity detunings $\Delta_c$ from red (a) to zero (b) to blue (c). A wide array of behaviors, including dynamical (in)stability depicted in (dashed) solid lines can be obtained. For an attractive interaction ($C_6 \le 0$) as considered here and at negative cavity detuning (Fig.~\ref{fig:SS_cases}(a)), optical multi-stability emerges as a result of the resonance crossing. While the middle branch with the negative slope is always unstable (dotted line) the two other branches' stabilities depend on the cavity detuning. When both the lower and upper branches are stable and the middle one is unstable, the cavity field follows a hysteretic behavior, i.e. by increasing the pump intensity, eventually, the lower branch mode abruptly jumps into the upper one when the lower branch ends. On the other hand, if the pump intensity is decreased, the field intensity decreases and jumps back down to the lower branch around this dynamically unstable region when the upper branch ends, similar to the response of a polariton ensemble with contact interaction~\cite{Carusotto2013}.

For a resonant excitation depicted in Fig.~\ref{fig:SS_cases}(b), only one branch exists, typically referred to as the \emph{pump-only} branch, whose stability depends on the pump intensity. As the photon density increases, the nonlinearity modifies the behavior and deviates from the quadratic trend. Since the detuning increases with the photon number, the effective pumping rate decreases which leads to a sub-linear growth of the cavity field intensity. 

For positive cavity detuning as shown in Fig.~\ref{fig:SS_cases}(c) there is only one branch as the resonant case. However, unlike Fig.~\ref{fig:SS_cases}(b) the cavity photon density grows monotonically with the pump intensity. This is because of the lower photon density compared to the resonant case to start with. The attractive potential detunes the cavity further as the photon density increases, and hence the collective nonlinearity remains low leading to a monotonic growth, almost quadratically.

In Fig.~\ref{fig:SS_cases}(d)-(i) we present the real (dark blue) and imaginary (light blue) parts of the Bogoliubov dispersion of points (A)-(F) denoted in panels (a)-(c), respectively. While the stability of the lower branch at point A can be deduced from the always-negative imaginary part in Fig.~\ref{fig:SS_cases}(d), the instability of the middle branch at point B can be deduced from the positive imaginary parts of the dispersion at $k = 0$ in Fig.~\ref{fig:SS_cases}(e).

At point C on the upper branch, depicted in Fig.~\ref{fig:SS_cases}(f), the two Bogoliubov branches cross in a finite momenta range away from $k= 0$ and exceptional points emerge. The branches are stuck in the vicinity of the crossing point and give rise to a flat region where $\text{Re}(\omega_\pm) = 0$. On the other hand, the imaginary parts split, and as soon as one of them turns positive, the system becomes dynamically unstable. Similar behavior can be observed under resonant excitation, hosting a finite region of MI (cf. Fig.~\ref{fig:SS_cases}(g)). At larger detunings $\Delta_c = 2$, the dispersion shows a roton minimum which transitions from stable behavior at weaker pump strengths (Fig.~\ref{fig:SS_cases}(h)) to a finite region of instability, or roton minimum softening, at stronger pump strengths (Fig.~\ref{fig:SS_cases}(i)). 

\begin{figure}
    \centering
    \includegraphics[width=8.8
    cm]{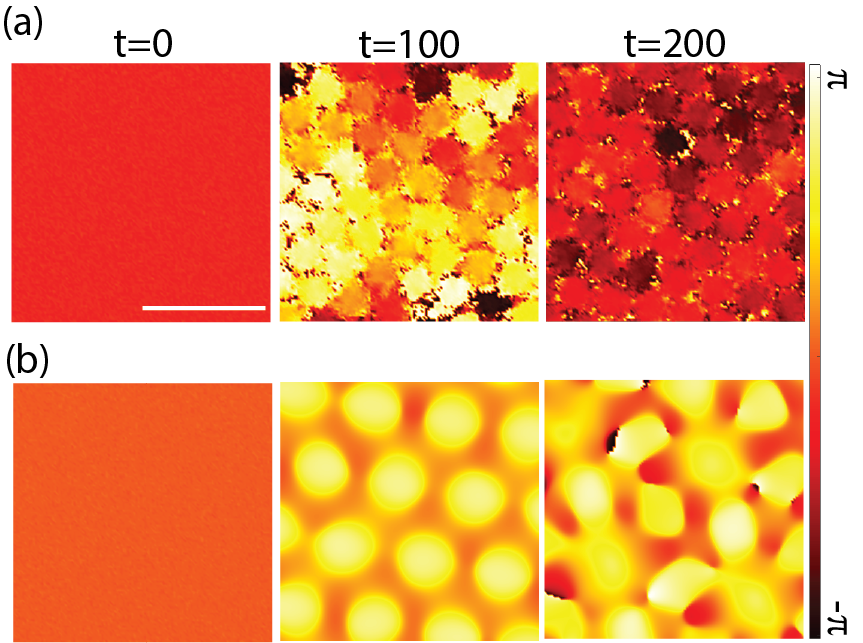}
    \caption{\textbf{Phase distribution and polariton flow in closed and open systems.} Snapshots of cavity field phase for (a) conservative polariton condensates and (b) the driven-dissipative Rydberg polariton ensemble under a coherent pump, illustrating the stability of the closed case and the ultimate disappearance of the patterns in the open system. The scale bar corresponds to $L = 10$.}
    \label{fig:open_pattern_phase}
\end{figure}

As discussed in the Supplementary Materials for the closed system, the MI at finite wavevectors implies the parametric gain which can lead to the spontaneous breaking of the translational symmetry and the emergence of the spatial patterns. To examine the ensemble's behavior in the presence of the MI, we time evolve the field dynamics given by the generalized GPE in Eq.~\ref{eq:GPE}. Figure~\ref{fig:schematics}(d) shows a few snapshots of the cavity field intensity at point F in Fig.~\ref{fig:SS_cases}(c). Similar to the conservative polariton dynamics depicted in Fig.~\ref{fig:schematics}(c), upon starting from a noisy initial state at $t = 0$, density patterns establish at longer times, e.g. $t = 100$. In stark contrast to the closed system with stable patterns, however, the patterns in the driven-dissipative polariton cloud are not stable at longer times, as depicted in the field snapshot at $t = 200$. To shed light on the pattern instability of the open system, we compare the phase distribution of the ensemble in Fig.~\ref{fig:open_pattern_phase}. As can be seen in Fig.~\ref{fig:open_pattern_phase}(a) for the closed case, the polariton cloud establishes a constant phase at equilibrium, corresponding to a uniform chemical potential and hence a vanishing flow. The out-of-equilibrium polariton's phase, on the other hand, is not uniform as depicted in Fig.~\ref{fig:open_pattern_phase}(b). Interference between the inherent phase of the symmetry-broken field and the phase of the external coherent drive leads to a non-vanishing flow towards the high-density points which limits the lifetime of the patterns (cf. Fig.~\ref{fig:open_pattern_phase}(b), the snapshot at $t = 200$).
%
%

To avoid this interference effect and stabilize the density patterns, we investigate the dynamics of the Rydberg polariton cloud subject to an incoherent pump. The driving term in the generalized nonlocal GPE of Eq.~\ref{eq:GPE} for an incoherent pump intensity $P(\mathbf{r})$, with phenomenological coupling and saturation parameters of $\gamma_R$ and $R$, can be described as~\cite{Wouters2007, Carusotto2013} 
\begin{equation}
    i\frac{R ~ P(\mathbf{r})}{2\gamma_R + 2R~|\mathcal{E}(\mathbf{r},t)|^2} \mathcal{E}(\mathbf{r},t)\, .
\end{equation} 
Unlike the coherently-driven case, this equation is U(1)-symmetric implying phase freedom of the polariton ensemble. Following a similar approach for the coherent drive, the fluctuation spectrum around the flattop solution $\psi_0$ can be determined as
\begin{equation*}
    \omega_\pm(\mathbf{k}) = -i \frac{\Gamma_\textrm{eff}(\mathbf{k})}{2} \pm \sqrt{\frac{k^2}{2} \left(\frac{k^2}{2} + 4\pi \psi_0^2 \Tilde{W}_R(\mathbf{k})\right) - \frac{\Gamma_\textrm{eff}(\mathbf{k})^2}{4}}\, ,
\end{equation*}    
where the effective gain/loss rate is defined as 
\begin{equation}
    \Gamma_\textrm{eff}(\mathbf{k}) = \left(\frac{R^2 P_0}{\left(\gamma_R + R|\psi_0|^2\right)^2}  - 4\pi  \Tilde{W}_I(\mathbf{k}) \right) \psi_{0}^{2}\, .
\end{equation}

Aside from $\Gamma_\textrm{eff}$, this dispersion is very similar to the conservative case as detailed in Eq.~9 (cf. Supplementary Materials), and hence similarities in density patterns and their stability are expected, as well.

\begin{figure}
    \centering
    \includegraphics[width=8.7
    cm]{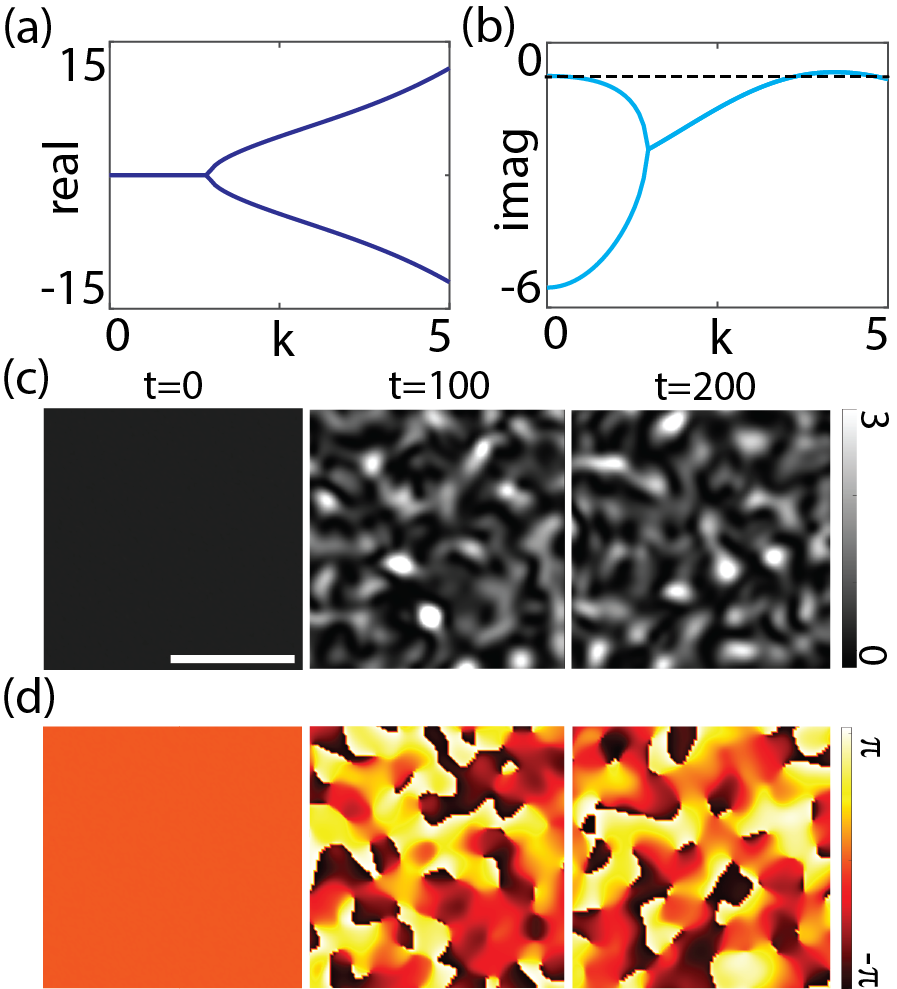}
    \caption{\textbf{Dispersion and pattern formation under an incoherent drive.} (a) Real and (b) imaginary parts of the Bogoliubov dispersion in an unstable flattop solution with $\Delta_\text{ex} = 1, \gamma = 0.1, \kappa = 0.5, C_6 = -1$. Snapshots of spatial (c) photon density $|\psi_0|^2$ and (d) its corresponding phase at various times. The scale bar corresponds to $L = 10$.}
    \label{fig:open_pattern_inc}
\end{figure}

Figure~\ref{fig:open_pattern_inc}(a) and (b) exemplify the real and imaginary parts of the Bogoliubov dispersion of an incoherently driven cavity, respectively. The dispersion features a finite range of exceptional-point momenta where the real parts of two branches coalesce and their imaginary parts depart from each other. Furthermore, as can be seen in Fig.~\ref{fig:open_pattern_inc}(b), there is a finite range of modulational instability where $\textrm{Im}(\omega(k))\ge 0$. To investigate the emergence of patterns due to MI and their stability, we time evolve the GPE in Eq.~\ref{eq:GPE} with an incoherent drive. The cavity field intensity and its corresponding phase at a few different times are depicted in Fig.~\ref{fig:open_pattern_inc}(c),(d), respectively. As can be seen, unlike the coherent drive, there is no particular phase pattern, and hence no flow of polaritons which would lead to pattern collapse, as highlighted in the field density profile at long times in Fig.~\ref{fig:open_pattern_inc}(c). 
%
%

\emph{Conclusion - } In this work, we present the first findings pertaining to the emergence of dynamically unstable phases in the novel platform of Rydberg exciton-polaritons featuring nonlocal and long-range interactions characterized by the intrinsic length scale of the potential, denoted as the blockade radius. We examined various steady-state phases, encompassing scenarios of multi-stability and modulational instability. Further, we elucidated the emergence of patterned phases and showed that such patterns are inhibited by coherent optical pumping. It is important to highlight that the spectral characteristics of the incoherently-driven ensemble closely resemble those of the conservative dynamics, suggesting a heightened level of robustness in behavior.

This novel platform of Rydberg polaritons offers a distinctive opportunity to investigate a unique regime characterized by strongly interacting, and out-of-equilibrium phases in many-body systems. This includes but is not limited to the study of quantum fluids supporting solitons and vortices~\cite{Lagoudakis2008, Klaus2022}, Bose-Hubbard models, quantum synchronization in the presence of long-range interactions~\cite{Moroney2021}, as well as the emergence of topological effects and vertices~\cite{Ma2016, Grass2018}. Furthermore, it would be interesting to extend the current study beyond the mean-field description of the cavity field to obtain further insights into the imprinted correlations between photons, e.g., by using the Quantum Monte Carlo (QMC) technique to incorporate fluctuations and subsequently examine photonic correlation effects~\cite{Cinti2014}. 

\section*{acknowledgment}
The authors would like to thank Thomas Pohl and Jens Hertkorn for their insightful discussions, and Furqan Hashmi for his contributions at the early stage of this project. HA acknowledges the Purdue University Startup fund and Purdue College of Science, the financial support from the Industry-University Cooperative Research Center Program at the US National Science Foundation under Grant No. 2224960, and the Air Force Office of Scientific Research under award number FA9550-23-1-0489.

\bibliography{ref}
\end{document}